\documentclass[preprint,review,12pt]{elsarticle}
\usepackage{amssymb}
\usepackage{bm}
\usepackage{ulem}
\usepackage{xcolor}
\usepackage{lineno}

\begin{document}

\begin{frontmatter}

\title{Numerical simulation of heat extraction by a coaxial ground heat exchanger under freezing conditions}
\author[a]{G.P. Vasilyev}
\author[b]{N.V. Peskov\corref{*}} 
\ead{peskov@cs.msu.ru}
\cortext[*]{Corresponding author}
\author[b]{T.M. Lysak} 
\address[a]{OAO “INSOLAR-INVEST”, Moscow, Russian Federation.}
\address[b]{Faculty of Computational Mathematics and Cybernetics, \\Lomonosov Moscow State University,  Moscow, Russian Federation.}

\begin{abstract}
A new finite-difference model of heat transfer inside a shallow coaxial ground heat exchanger and in the surrounding layered soil is presented, taking into account the freezing of ground moisture. Three modes of heat exchanger operation are numerically simulated: stationary mode, transient mode and controlled mode. In the stationary mode, estimates of the sensitivity of the heat carrier fluid outlet temperature to changes in the heat exchanger parameters are calculated. In all modes, close attention is paid to demonstrating the difference in the results at a negative temperature of the fluid, calculated with and without taking into account the freezing of ground moisture. It is shown that this difference, caused by the zero-curtain effect, can range from 10\% in the stationary mode to 35\% in the control mode.

\end{abstract}


\end{frontmatter}

\section{Introduction}

Ground source heat pump (GSHP) is a highly efficient renewable energy technology which allows sufficient reduction in energy consumption for building heating and cooling \cite{Menegazzo_2022}. An important part of each GSHP system is a ground heat exchanger (GHE) installed in a vertical borehole or horizontal trench. Thermal energy is extracted from the ground or absorbed by the ground due to the effective heat exchange between the fluid,  which circulates in the GHE, and the adjacent soil. The effective design of GHEs requires a good understanding of the thermal processes in the system GHE-surrounding soil. 

Three main approaches to the investigation of the system GHE-surrounding soil can be distinguished: analytical, numerical and experimental. Analytical approach allows obtaining analytical dependencies under a number of strong simplifications about the thermal processes. More complicated 2D or 3D geometry   models require numerical simulation, but use the results of analytical approach and experimental approaches for their testing. A lot of information on the modeling of thermal processes for various types of GHE can be found in reviews \cite{Zhao_2022, Cho_2013}. In \cite{Cho_2013}, a list of commonly used software packages for GSHP design can be found as well. 

While most models focus on normal GHE operation with above-zero inlet temperature, the effect of phase change on GHE operation is of great importance in cold regions, where the circulating fluid inlet temperature must be below zero to extract the required amount of heat. Freezing-thawing processes around the GHE change the thermal characteristics of the soil thus affecting the thermal exchange between the pipe fluid and the surrounding soil. In particular, a number of experiments show that freezing increases the heat transfer rate and the coefficient of performance \cite{Cao_2018, Eslami_2012, Yang_2014, Meng_2020, Tu_2017, Wu_2020, Zhang_2021_2}. As a result, utilization of latent heat from the groundwater freezing allows, for example, the reduction of vertical GHE length for the GSHP with the same characteristics \cite{Eslami_2012}.  

Another consequence is the possible damage of the pipe due to the ice lens formation in the grout \cite{Erol_2015} and the impact of freeze-thaw cycles on GHE hydraulic conductivity \cite{Kupfernagel_2021, Kirshbaum_2018, Anbergen_2014, Anbergen_2015}.  In \cite{Kirshbaum_2018}, the influence of freeze-thaw cycles on aging effect of U-shaped GHEs were studied using the large-scale test rig determining the hydraulic conductivity of GHE systems. In \cite{Kupfernagel_2021}, a pilot-scale experiment was built to test a 1-m section of a typical GHE under freezing-thawing conditions.  Freeze-thaw cycles seem to be a leading cause for a potential violation of the system’s integrity \cite{Kupfernagel_2021, Anbergen_2014}. In \cite{Anbergen_2015}, the thermal processes in the saturated porous medium around the GHE were simulated using FEFLOW 2D axisymmetric model with C++ plug-in for  phase changes between solid and liquid phases. Possible ways to avoid GHE pipes deformation and reduce flow resistance under freezing conditions were studied via experimental tests in \cite{Wang_2013} and using numerical simulation in \cite{Wang_2018}. 

Note that the damage of the pipes is mainly studied by experimental tests. On the contrary, the role of numerical simulation in studying the efficiency of thermal transfer processes  in the system GHE-surrounding soil is very large. Up to now a number of math models for this aim  was developed, which take into account freezing-thawing processes in the soil around the vertical and horizontal GHEs. The major part of all math models is the heat transfer equation for the surrounding soil. The freezing-thawing processes are described  in the model using a latent heat term in the heat transfer equation and  the heat transfer model \cite{Yang_2014}  with effective heat capacity to describe the three phases -- solid, liquid and mushy. This method allows avoiding one of the difficulties in numerical simulation for phase change: how to treat the movable interface between the phases. The other important  part of math model is temperature distribution inside the GHE pipe. Some investigations do not consider this temperature distribution, operating with the total heat gain/loss in GHE \cite{Yang_2014, Santa_2019}. The others (see \cite{Zheng_2016}, for example ) use  a dual continuum approach \cite{Diersch_2011}. Under such approach, the soil is described in 3D geometry, while the two branches of  U-shaped GHE are described in 1D geometry. To simplify the description of the temperature distribution inside the U-shaped pipe, the pipe is replaced by a mono-tube with equivalent diameter \cite{Yang_2013}.

Without a doubt, detailed 3D or even 2D calculation of thermal processes in the GHE and surrounding soil can provide accurate results. But such calculations are in most cases based on commercial software packages, such as FEFLOW, COMSOL, ABAQUS, OpenGeoSys , CFD ANSYS FLUENT and so on. Commercial software packages are also very useful at modeling the arrays of GHEs \cite{Meng_2020, Zhang_2021, Li_2019}. An obvious drawback of these packages is the impossibility of  numerical modeling of phase transfer processes in porous soil.  To overcome this drawback, some authors use C++ or FORTRAN plug-ins \cite{Anbergen_2015, Wang_2018} to commercial software package for  modeling phase transition between solid and liquid phases in porous surrounding soil. Another disadvantage of detailed 3D simulation is the large time of calculation and large computational resources necessary for the long-term simulation of the system GHE-surrounding soil with phase change processes.

To reduce the computational resources, simplified 1D and 2D model for vertical GHEs were developed. For example in \cite{Tu_2019}, an improved thermal resistance and capacity model (RC) was proposed for the heat transfer modeling between vertical single U-tube GHEs and the frozen soil. The model was verified through experiment tests  and  numerical simulation on basis of 3D  CFD ANSYS  model. In \cite{Yang_2014}, a 2D model for temperature field modeling around the GHE was developed. The model is based on heat transfer equation in cylindrical geometry with a source term and Dirichlet  boundary conditions to represent the GHE. In \cite{Eslami_2012}, 1D radial numerical heat transfer model is developed to evaluate heat transfer from the borehole wall to the ground with taking into account three phases: ice-soil, water-soil and transition phase. For the boundary conditions at the borehole wall and the outer boundary of computational area the temperature, measured experimentally, was used.  Another approach to reduce the computational time was demonstrated in \cite{Santa_2019} for the modeling of a GHE array applied to an existing historical building in Venice (Northern Italy). In this paper, two different scale models were developed. The results of the coarser large scale model were used as initial and boundary conditions for the fully discretized small scale model, developed for the detailed description of freezing-thawing processes in the close vicinity of the GHE. 

The processes of freezing-thawing are closely related to the processes of groundwater seepage and flow.  The influence of these two factors -- freezing-thawing together with  ground water seepage, are taken into account in \cite{Meng_2020, Zhang_2021_2, Zhang_2021}  in modeling the temperature field in the surrounding soil for the array of vertical U-shaped GHEs. The temperature field equation for the porous surrounding soil with convection and ice/water phase and  Brinkman equation for water transport in porous media are the basis of the 3D model, which is solved  by the COMSOL software. The model does not describe the temperature inside the pipes. Instead it uses the iteration processes to determine the inlet and output temperature on basis of the soil model. The similar model was developed in \cite{Li_2019} for artificial freezing. 

This paper proposes a combined mathematical model containing a 1D model of heat transfer in a fluid circulating inside a vertical coaxial GHE and a 2D model of heat conduction in a cylindrically symmetric surrounding soil. The model takes into account temperature changes in the inner and annular pipes, soil freezing processes, geothermal temperature gradient and horizontal soil stratification. Numerical implementation of the model, based on the finite difference method and the MatLab {\bf ode15s} numerical solver, makes it possible to investigate the effect of soil freezing on the outlet temperature of the coaxial heat exchanger.

Coaxial GHE have a number of advantages in  comparison with U-shaped one, especially in the case of deep boreholes (1000 to 3000 meters depth). Among them is a simple installation procedure,  moderate temperature difference between the secondary fluid and the surrounding ground,  the possibility to use water as a secondary fluid even in colder countries and to ignore its local thermal resistance in comparison with U-shaped GHE  \cite{Acuna_2010, Li_2020}. The comparison of different types of GHE can be found in \cite{Zarrela_2017}, where the equivalent ground thermal conductivity  was evaluated by Thermal Response Test. A  review on various approaches for investigation of coaxial GHEs can be found, for example, in \cite{Wang_2021}.

Geothermal gradient was taken into account in an analytical model \cite{Luo_2019}  and  semi-analytical models \cite{Wang_2021, Wang_2022} proposed for shallow (100 to 200 m depth, \cite{Wang_2021}, \cite{Luo_2019}) and deep (1000 to 3000 meters depth, \cite{Wang_2022}) coaxial BHE.

 The vertical geological structure of the adjacent ground was considered in \cite{Liu_2019} at numerical modeling of a  deep coaxial GHE. 

However, to our knowledge, freezing-thawing processes have not yet  been considered for coaxial GHE. It should be noted that, in deep GHEs, only pure water (without anti-freeze) with positive inlet temperature is allowed to use to prevent pollution on deep groundwater resources \cite{Cai_2022}. Therefore, in our paper, we focus on a shallow vertical coaxial GHE with negative inlet temperature of circulating fluid.

\section{Formulation of the problem}

\subsection{Scheme of coaxial ground heat exchanger}

\begin{figure}[h]
\centering
\includegraphics{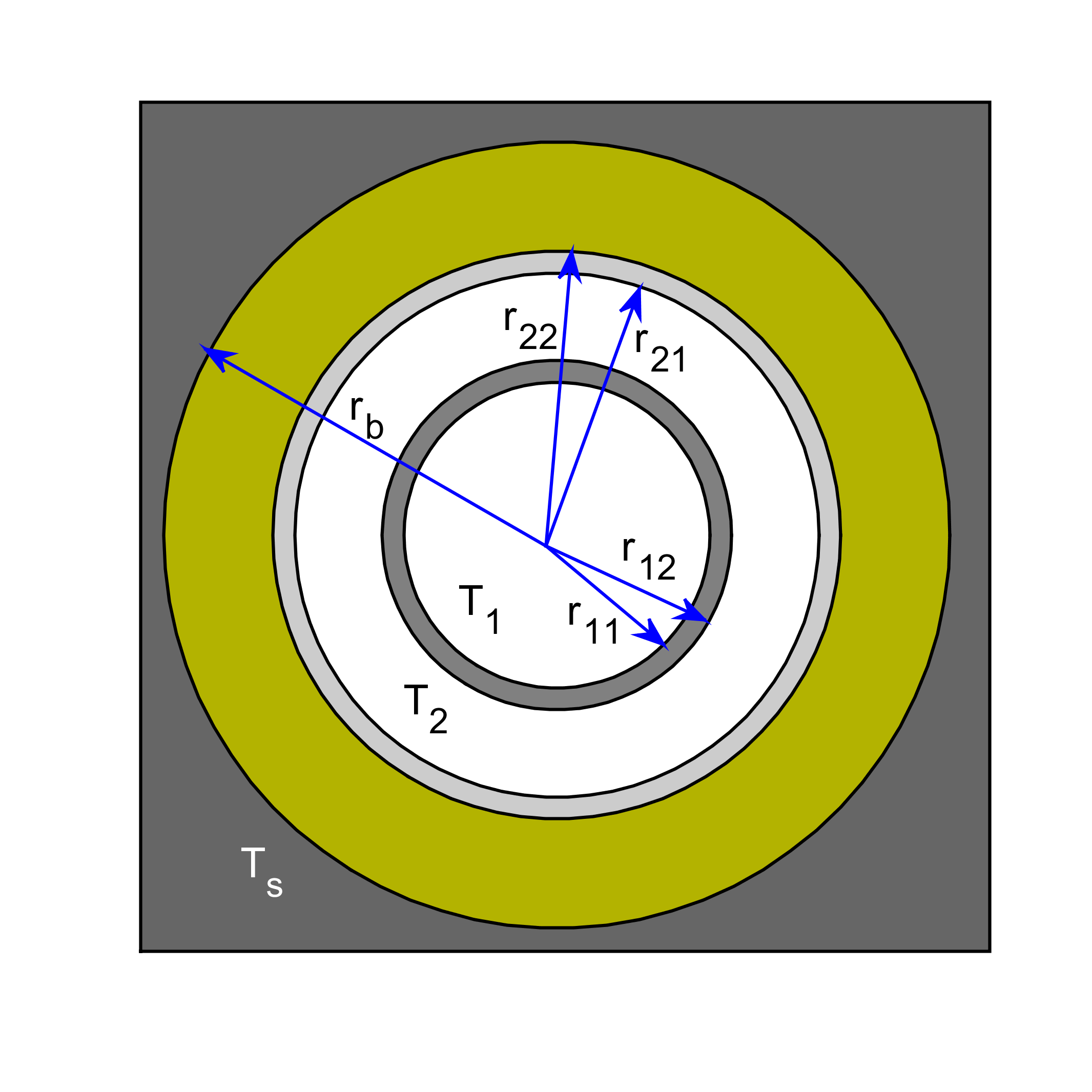}
\caption{Out of scale scheme of coaxial ground heat exchanger.}
\label{she}
\end{figure}

We imagine a heat exchanger in the form of two coaxial pipes located in a vertical borehole and fixed in it with the help of a grout. The cross section of the heat exchanger is shown schematically, not to scale, in the Figure \ref{she}. 

The cross section of the inner pipe is shown in the center of the figure, $r_{11}$ and $r_{12}$ are the inner and the outer radius of this pipe, respectively. Coaxial with the inner pipe is the outer pipe, its inner and outer radius are designated $r_{21}$ and $r_{22}$, respectively. The radius of the borehole is denoted by $r_b$, the space between the outer pipe and the soil outside the borehole is filled with grout. In the heat exchanger scheme, we take into account the thickness of the pipe walls and the thickness of the grout layer to estimate the magnitude of heat flows between the pipes and between the fluid and the soil.

In the process of heat exchange, the heat-carrying fluid moves along the inner tube and in the annular space between the tubes in opposite directions. We will consider a heat exchanger in which the fluid moves from top to bottom along the annular area between the pipes and rises up the inner pipe. Such a mode is usually used to extract heat from the ground. The temperature of the fluid in the inner pipe will be denoted by $T_1$, in the space between the pipes -- by $T_2$, and through $T_s$ we will denote the temperature of the soil. In the general case, the temperatures $T_1$, $T_2$, and $T_s$ are unknown functions of time and spatial coordinates.

\subsection{Mathematical statement of the problem}

To calculate the temperature of the fluid and the soil surrounding the borehole, we will solve the initial-boundary value problem in a circular cylinder of radius $r_d$ and height $H$ with a heat exchanger located along the vertical axis of the cylinder. The problem will be solved in a cylindrical coordinate system with the origin located in the center of the upper end of the inner pipe and the axis $z$ directed downward and coinciding with the axis of the inner pipe. We assume that the temperature of the fluid $T_1$ and $T_2$ can depend only on the time $t$ and the coordinate $z$, $T_1(t,z)$ and $T_2(t,z)$. We do not take into account the thermal conductivity of the fluid, assuming that heat in the exchanger is distributed only due to the flow of the fluid and heat transfer through the walls of pipes and borehole. 

Regarding the soil, we assume that its physical properties may depend on the coordinates $r$ and $z$, but do not depend on the azimuth angle, $T_s=T_s(t,r,z)$. The side wall of the cylinder at $r = r_d$ is considered to be thermally insulated, and a constant temperature is maintained on the bases of the cylinder. Moreover, we will take into account the geothermal gradient, which is essential for long heat exchangers. Thus, we will not take into account seasonal changes in temperature on the surface of the earth, as well as changes in the properties of the surface itself. It can be assumed that the upper part of the heat exchanger is located at a depth of about 10 m under the earth surface, where the soil temperature is practically constant throughout the year and is equal to the average annual air temperature on the surface.

Based on the assumptions made, the problem of calculating the temperature of the fluid and the soil can be written in the following form.

\begin{eqnarray}
&&c_f\rho_fa_1\frac{\partial T_1(t,z)}{\partial t} = 
c_fv_f\frac{\partial T_1(t,z)}{\partial z} - q_1(T_1,T_2), \\
&&c_f\rho_fa_2\frac{\partial T_2(t,z)}{\partial t} = 
-c_fv_f\frac{\partial T_2(t,z)}{\partial z} + q_1(T_1,T_2) - q_2(T_2,T_{sb}), \\
&&c_s\rho_s\frac{\partial T_s(t,r,z)}{\partial t}  = 
\frac{1}{r}\frac{\partial}{\partial r}\left[r\lambda_s\frac{\partial T_s(t,r,z)}{\partial r}\right] + \frac{\partial}{\partial z}\left[\lambda_s\frac{\partial T_s(t,r,z)}{\partial z}\right].
\end{eqnarray} 

\noindent
Here are the constant parameters of the fluid: $c_f$, [J/(kg$\cdot$K)], is the heat capacity; $\rho_f$, [kg/m$^3$], is the density; and $v_f$, [kg/s], is the mass velocity. The cross section area of the inner pipe and the annular region between the pipes are denotes as $a_1=\pi r_{11}^2$ and $a_2=\pi r_{21}^2- \pi r_{12}^2$. In contrast to the fluid parameters, the soil parameters $c_s$ and $\rho_s$, as well as the thermal conductivity of the soil, $\lambda_s$ [W/(m$\cdot$K)], in the general case, can depend on spatial coordinates and on temperature $T_s$. However, in order not to complicate the formulation of the problem and the numerical solution, we will consider the smooth dependence of the soil parameters on the coordinates.

The terms $q_1$ and $q_2$, [W/m], describe heat transfer between fluids flowing down and upwards, and between fluid and soil, respectively. As usual, we calculate the values of these terms from the boundary value problem for the stationary heat equation in the rings $r_{11}<r<r_{12}$ and $r_{21}<r<r_b$ with Robin boundary conditions at the pipe walls and the Dirichlet condition at the borehole wall. From the solution of these problems,

\begin{eqnarray}
&&q_1(T_1,T_2) = \frac{2\pi\lambda_1(T_1-T_2)}{\ln\left(\frac{r_{12}}{r_{11}}\right) + 
\frac{\lambda_1}{\alpha_{11}r_{11}}
+ \frac{\lambda_1}{\alpha_{12}r_{12}}}, \\
&&q_2(T_2,T_{sb}) = \frac{2\pi\lambda_2(T_2-T_{sb})}{\ln\left(\frac{r_{22}}{r_{21}}\right) + \frac{\lambda_2}{\lambda_g}\ln\left(\frac{r_b}{r_{22}}\right) + 
\frac{\lambda_2}{\alpha_{21}r_{21}}}.
\end{eqnarray}

\noindent
Here $\lambda_1,\, \lambda_2,\, \lambda_g$ are the thermal conductivity coefficients of the walls of the inner and outer pipes and the grout, respectively; $\alpha_{11},\,\alpha_{12},\, \alpha_{21}$ are the coefficients of convective heat transfer between the fluid and pipe walls: $T_{sb}$ is the temperature of the soil adjacent to the borehole wall.

Note that the equations,  similar to equations (1)-(5), were written in  \cite{Liu_2019}  for a deep coaxial GHE,  without taking into account the grout and the freezing-thawing processes.

The equations (1), (2) are defined in segment $0<z<H$ with the boundary conditions
\begin{equation}
T_1(t,H)=T_2(t,H);\; T_2(t,0) = T_{in},
\end{equation}
where $T_{in}$ is the fluid temperature at the heat exchanger inlet. In addition, we will need the outlet temperature $T_{out}(t)=T_1(t,0)$, as well as the heat extraction rate
\begin{equation}
q_{ex}(t) = c_fv_f(T_{out}(t)-T_{in}).
\end{equation}

The equation (3) is defined in the rectangle $\{r_b<r<r_d,\, 0<z<H\}$. In accordance with the assumptions made above, we set the boundary conditions for Eq. (2) as follows
\begin{eqnarray}
&& \lambda_sr_b\frac{\partial T_s(t,r,z)}{\partial r}\bigg|_{r=r_b} = q_2,\; 
\frac{\partial T_s(t,r,z)}{\partial r}\bigg|_{r=r_d} = 0, \\
&& T_s(t,r,z)\big|_{z = 0} = T_{top},\; T_s(t,r,z)\big|_{z = H} = T_{bot},
\end{eqnarray}
where $T_{top}$ is the specified soil temperature at the top of the heat exchanger (assumed to be equal to the average annual temperature of the ambient air), and $T_{bot}$ is calculated by the equation
\[T_{bot} = T_{top}+G_tH,\]
where $G_t$ is the geothermal gradient

As the initial conditions for equations (1)-(3) we put
\begin{equation}
\label{tf0}
T_1(0,z)=T_2(0,z)=T_{top}+G_tz,
\end{equation}
and also 
\begin{equation}
\label{ts0}
T_s(0,r,z)=T_{top}+G_tz,
\end{equation}

\subsection{Numerical scheme}

We will solve problem (1)-(11) numerically using the finite difference method. For this purpose, we introduce a uniform grid along the $z$ coordinate, $z_i=ih$, $i = 0,1,\dots,N_z$, $h = H/N_z$; and an exponential grid along the $r$ coordinate,
\[r_j = r_b\exp\left[\frac{j-0.5}{N_r}\ln\left(\frac{r_d}{r_b}\right)\right],\,
j = 1,2,\dots,N_r.\]
An exponential grid has been used in several papers cited above and is needed to record the rapid change in soil temperature near borehole. Obviously, it is inspired by the fundamental solution of the Laplace equation in polar coordinates, $T(r)\sim \ln r$.

Passing to grid functions, $T^1_i(t) = T_1(t,z_i)$, $T^2_i(t) = T_2(t,z_i)$, $T^s_{i,j}(t) = T_s(t,r_j,z_i)$; and replacing spatial derivatives in equations (1)-(3) with finite differences, we obtain a system of $N_z(N_r+2)$ ordinary differential equations.
\begin{eqnarray}
\label{et1}
&&c_f\rho_fa_1\frac{dT^1_i}{dt} = c_fv_f\frac{T^1_{i+1}-T^1_i}{h}-q_1(T^1_i,T^2_i), \\
&&i = 0,1,\dots,N_z-1,\, T^1_{N_z} = T^2_{N_z}; \nonumber \\
\label{et2}
&&c_f\rho_fa_2\frac{dT^2_i}{dt} = -c_fv_f\frac{T^2_i-T^2_{i-1}}{h}+q_1(T^1_i,T^2_i)
-q_2(T^2_i,T^s_{i,1}), \\
&&i =1,2,\dots,N_z,\, T^2_0 = T_{in}; \nonumber \\
\label{ets}
&&(c_s\rho_s)_{i,j}S_j\frac{dT^s_{i,j}}{dt} = R_{i,j}+Z_{i,j},\, i = 1,2,...,N_z, \, j = 1,2,...,N_r;
\end{eqnarray}
where $S_j = \pi(r^{*2}_j-r^{*2}_{j-1})$. $r^*_0 = r_b$, $r^*_j = 0.5(r_j+r_{j+1})$.
\begin{eqnarray}
&&R_{i,1} = q_2(T^2_i,T^s_{i,1}) - 
2\pi r^*_1\lambda_{i,1}^r\frac{T^s_{i,1}-T^s_{i,2}}{r_2-r_1}; \nonumber\\
&&R_{i,j} = 
2\pi r^*_{j-1}\lambda_{i,j-1}^r\frac{T^s_{i,j-1}-T^s_{i,j}}{r_j-r_{j-1}} -
2\pi r^*_j\lambda_{i,j}^r\frac{T^s_{i,j}-T^s_{i,j+1}}{r_{j+1}-r_j},\\
&&j = 2,3,...,N_r-1; \nonumber \\
&&R_{i,N_r} = 
2\pi r^*_{N_r-1}\lambda_{i,N_r-1}^r\frac{T^s_{i,N_r-1}-T^s_{i,N_r}}{r_{N_r}-r_{N_r-1}}, \nonumber\\
&&i = 1,2,...,N_z, \nonumber
\end{eqnarray}
where
\[\lambda_{i,j}^r = \frac{2\lambda_{i,j}\lambda_{i,j+1}}{\lambda_{i,j}+\lambda_{i,j+1}}.\]
\begin{eqnarray}
&&Z_{i,j} = \frac{S_j}{h^2}\left[\lambda_{i-1,j}^z(T^s_{i-1,j}-T^s_{i,j}) -
\lambda_{i,j}^z(T^s_{i,j}-T^s_{i+1,j})\right], \\
&&i = 1,2,...,N_z,\, j = 1,2,...,N_r; \nonumber
\end{eqnarray}
where
\[\lambda_{i,j}^z = \frac{2\lambda_{i,j}\lambda_{i+1,j}}{\lambda_{i,j}+\lambda_{i+1,j}}.\; T^s_{0,j} = T_{top},\, T^s_{N_z+1,j} = T_{bot}. \]

We have written finite-difference equations for the general case when soil parameters depend on spatial coordinates. Obviously, for a homogeneous soil, the equations are somewhat simplified. 


\section{Model parameters}

For simulations that demonstrate the capabilities of the model, it is necessary to set the values of the parameters of the heat exchanger, heat-carrying fluid and ground. The parameters of shallow ground heat exchangers can take values in fairly wide ranges \cite{Shah_2022}. In this work, we do not consider any special heat exchanger. Therefore, for numerical simulation, we have chosen the typical values of the parameters that are found in the literature. These values are presented in Table 1.
  
\begin{table}[h]
\caption{Heat exchanger parameters}
\centering
\begin{tabular}{l l c}
\hline
     & Description  & Value \\
 \hline
$H$ & Pipes length & 100 m \\
$r_{11}$ & Inner radius of inner pipe & 0.04 m \\
$r_{12}$ & Outer radius of inner pipe & 0.043 m \\
$r_{21}$ & Inner radius of outer pipe & 0.074 m \\
$r_{22}$ & Outer radius of outer pipe & 0.08 m \\
$r_b$ & Radius borehole & 0.1 m \\
\hline
$\lambda_1$ & Thermal conductivity of inner pipe & 0.4 W/(m$\cdot$K) \\ 
$\lambda_2$ & Thermal conductivity of outer pipe & 40 W/(m$\cdot$K) \\ 
$\lambda_g$ & Thermal conductivity of grout & 1.5 W/(m$\cdot$K) \\ 
$\alpha_{11}$ & Convective coefficient at $r_{11}$ & 1000 W/(m$^2\cdot$K) \\ 
$\alpha_{12}$ & Convective coefficient at $r_{12}$ & 500 W/(m$^2\cdot$K) \\ 
$\alpha_{21}$ & Convective coefficient at $r_{21}$ & 500 W/(m$^2\cdot$K) \\ 
\hline
$c_f$ & Fluid heat capacity & 4500 J/(kg$\cdot$K) \\
$\rho_f$ & Fluid density & 1000 kg/m$^3$ \\
$v_f$ & Fluid mass velocity & 0.3 kg/s \\
\hline
\end{tabular}
\end{table}

Difference equations (12)-(16) make it possible to calculate the temperature distribution and heat fluxes in inhomogeneous soil. To demonstrate this possibility in this work, we have chosen an imaginary soil composed of several horizontal layers of minerals with different physical properties. The composition of the soil and the location of minerals in depth are presented in Table 2. The properties of minerals are described by the parameters, $\rho_s^0$, $c_s^0$, $\lambda_s^0$ -- density, heat capacity and thermal conductivity of dry matter, and volumetric moisture content in the substance of the layer, $m$. The bottom part of the table shows the thermophysical parameters for water and ice. In the next section, the parameters of water are denoted by the subscript $w$ ($\rho_w$, $c_w$, $\lambda_w$), and of ice -- by the subscript $ice$ ($\rho_{ice}$, $c_{ice}$, $\lambda_{ice}$).

\begin{table}[h]
\caption{Strati-graphic column of borehole}
\centering
\begin{tabular}{l c c c c c}
\hline
 Soil type  & $z_t-z_b$ & $\rho_s^0$ & $c_s^0$ & $\lambda_s^0$ & $m$ \\
 & m & kg/m$^3$ & J/(m$\cdot$K) & W/(m$\cdot$K) & m$^3$/m$^3$ \\
 \hline
clay & 0 - 5 & 1700 & 920 & 1.1 & 0.2 \\
limestone & 5 - 60 & 2500 & 840 & 1.0 & 0.15 \\
mudstone & 60 - 90 & 2600 & 800 & 1.8 & 0.1 \\
granite & 90 - 100 & 2700 & 790 & 1.1 & 0.05 \\
\hline
Average & 0 - 100 & 2510 & 827 & 1.25 & 0.13 \\
\hline
\hline
water & & 997 & 4200 & 0.57 & \\
ice & & 919 & 2108 & 2.25 & \\
\hline
\end{tabular}
\end{table}

\section{Simulation results}

Equations (\ref{et1})-(\ref{ets}) can be written in matrix form as
\begin{equation}
\label{mat}
\frac{d{\bm T}}{dt} ={\bf  A} {\bm T} + {\bm R},
\end{equation}
where $\bm T$ is a column vector containing $N_e=(N_r+2)N_z$ components.
\[T_i=T^1_i,\; T_{N_z+i}=T^2_i,\; i = 1,\dots,N_z;\]
 and
 \[\{T_{(j+1)N_z+i}=T^s_{i,j},\; i = 1,\dots,N_z\},\; j=1,\dots,N_r.\]
 The matrix $\bf A$ depends on the fluid and soil parameters, while the vector $\bm R$ takes into account the boundary conditions of the problem.The elements of the matrix $\bf A$ and the column vector $\bm R$ are easily determined from equations (\ref{et1})-(\ref{ets}).
 
 The matrix $\bf A$ contains $N_e^2$ elements, but it is essentially 5-diagonal. Namely, only $11N_z + (N_r-1)(5N_z-2)-5$ of them are nonzero. With the values of $N_z$ and $N_r$ taken in our calculations, the ratio of the number of non-zero elements to the total number of elements is of the order of $10^{-4}$. Thus, the matrix $\bf A$ is very strongly sparse. In this case, when solving system (\ref{mat}), it is very efficient to use the technique of sparse matrices, which makes it possible to radically reduce the size of RAM and the computation time. In this work, the solution of system (\ref{mat}) was carried out in MatLab using the {\bf ode15s} function, which effectively uses the sparseness of the matrix and overcomes the stiffness of the problem.
 
During the operation of the heat exchanger, the temperature of the fluid and the surrounding soil can drop below the freezing temperature of the ground moisture, $T_f$ (for definiteness, we assume $T_f=0^\circ$C). Antifreeze can be used as a heat carried fluid, but ground moisture will freeze at low temperatures, and then, when the temperature rises, melt. Thus, in ground moisture, a water-ice phase transition will occur. A change in the state of aggregation of moisture leads to a change in the physical properties of the soil and is accompanied by the release or absorption of the latent heat of the phase transition, $L$ = 334 kJ/kg. Therefore, adequate modeling of heat exchangers operating at the phase transition temperature must take this transition into account. However, due to computational difficulties, it is often ignored. In this work, for comparison, we present the results of calculations obtained both with and without taking into account the freezing-thawing of ground moisture.

In calculations that do not take into account the freezing of moisture, we assume that moisture behaves like antifreeze. In the following, a model with this assumption will be referred to as {\it model A}. In model A, at any temperature, the ground parameters in any stratum are calculated by the formula
\begin{equation}
\label{psp}
p_s = (1-m)p^0_s + mp_w,
\end{equation}
where $p$ denotes $\rho$, $c$ or $\lambda$.

To model the freezing-thawing of ground moisture, we use the method of apparent heat capacity \cite{Hashemi_1967, Hu_1996}. In this method, it is assumed that the water-ice transition occurs in a fairly narrow temperature range, $T_f-\Delta_f<T<T_f+\Delta_f$. When the temperature $T$ lies in this interval, then it is assumed that water and ice are simultaneously present in the soil. Moreover, the proportion of water, $w$, in the mixture is calculated by the formula
\begin{equation}
\label{wmix}
w = \frac{1}{2}\left(\frac{T}{\Delta_f}+1\right).
\end{equation}
And the ground parameters are determined by the equations
\begin{eqnarray}
\label{fsp}
&& \rho_s = (1-m)\rho^0_s +m[w\rho_w+(1-w)\rho_{ice}], \nonumber \\
&& c_s = (1-m)c^0_s +m[wc_w+(1-w)c_{ice}+L/(2\Delta_f)], \\
&& \lambda_s = (1-m)\lambda^0_s + m[w\lambda_w+(1-w)\lambda_{ice}]. \nonumber
\end{eqnarray}

At $T > T_f+\Delta_f$ the soil parameters are calculated by Eq. \ref{psp}. While at $T < T_f-\Delta_f$, when all the moisture in the soil is frozen, the soil parameters are calculated by the equation
\begin{equation}
\label{msp}
p_s = (1-m)p^0_s + mp_{ice},
\end{equation}

Allowance for moisture freezing makes system (\ref{mat}) non-linear and significantly increases the computation time. When solving (\ref{mat}) numerically, the matrix elements are recalculated only in the vicinity of the phase transition band. As $\Delta_f$ decreases, the number of elements to be updated decreases too, but the stiffness of the problem increases and the computation time increases. Therefore, the value of $\Delta_f$ is chosen so as to balance these opposite tendencies. In the calculations presented below, it is assumed that $\Delta_f = 0.1^\circ$C. In the following, the model that takes into account the freezing of soil moisture will be referred to as the {\it model F}.

\subsection{Stationary mode}

Under unchanged boundary conditions, the solution ${\bm T}(t,r,z)$ of system (\ref{mat}) at $t\to\infty$ asymptotically approaches the stationary solution ${\bm T}_{st}(r,z)$, which satisfies the system of algebraic equations
\begin{equation}
\label{mats}
{\bf A}{\bm T}_{st} + {\bm R} = 0.
\end{equation}

Figure 2 shows graphs of the stationary temperature of the fluid and the soil adjacent to the borehole wall, calculated both with allowance for moisture freezing (solid lines) and without freezing (dotted lines).

\begin{figure}[h]
\centering
\includegraphics{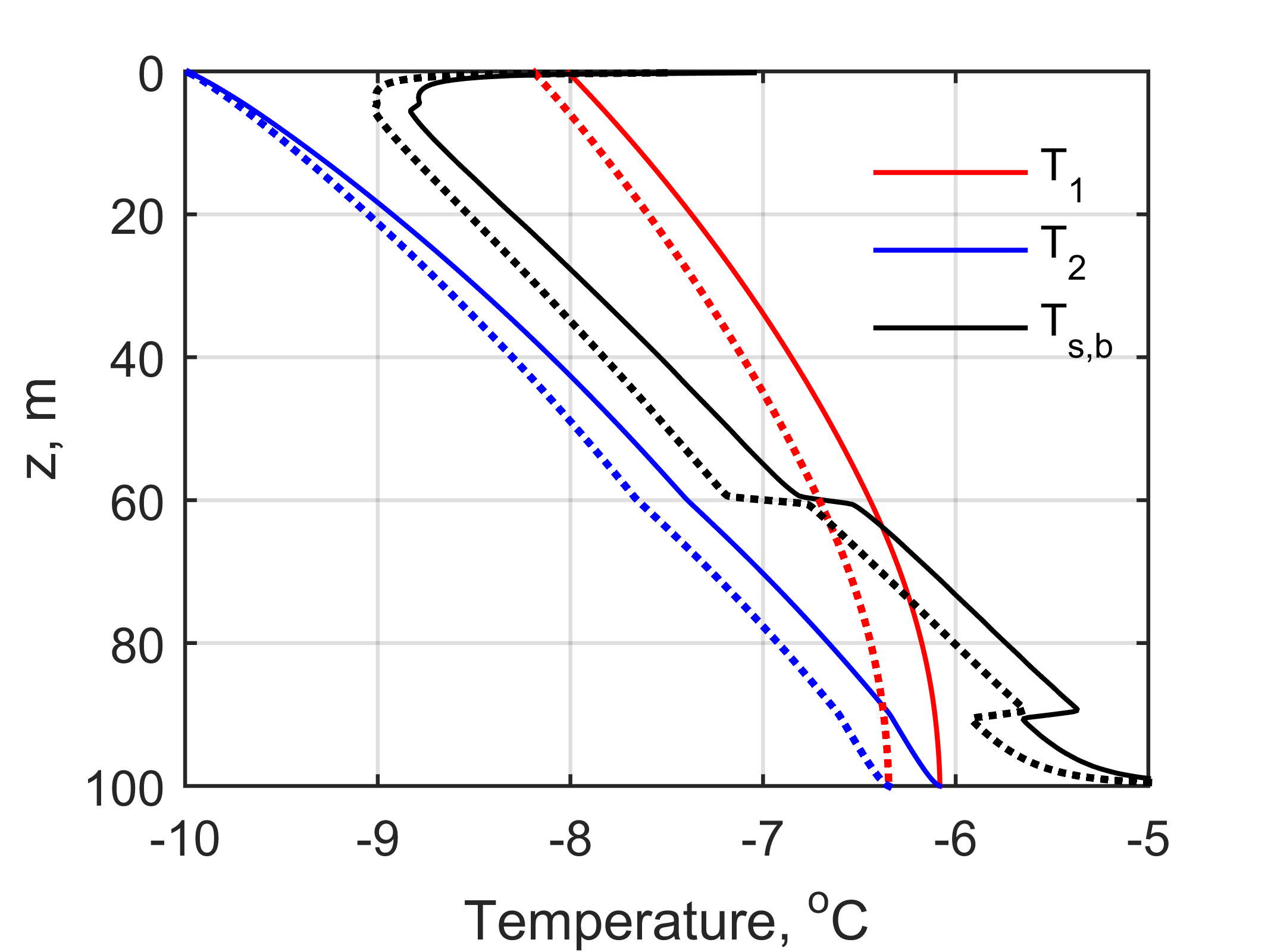}
\caption{Temperature of incoming, $T_2(z)$, outgoing, $T_1(z)$, fluid and adjacent soil, $T_{s,b}(z)$, in stationary mode at $T_{in} = -10^o$C, $v_f$ = 0.3 kg/s. Solid lines -- model F, dotted lines -- model A.}
\label{statb1}
\end{figure}

Freezing of ground moisture in our model leads to an increase in the thermal conductivity of the soil, and hence to an increase in the heat flux from the external soil to the heat exchanger. Therefore, in Fig. 2 temperature graphs for the model with freezing are shifted to the right relative to the graphs for the model without freezing. In particular, under given conditions, the temperature of the fluid at the outlet of the heat exchanger $T_{out} = -8.2^\circ$C in the model without freezing and $-8.0$ with freezing. Under considered conditions, a difference of 0.2 degrees in temperature means a difference of 972 kJ/h in heat extraction rate or 10 percent of the heat extraction rate in model A.

On the graphs of the temperature of the adjacent soil, a protrusion is distinguished at a depth of 60 to 90 m. This protrusion corresponds to a layer in which the thermal conductivity coefficient is significantly higher than in neighboring layers. Therefore, the heat flux from the outer soil in this layer is greater than in the neighboring ones, and, accordingly, the soil temperature is higher.

In our model and in many other models of heat exchangers, it is assumed that the heat flow between the coolant and the adjacent soil is proportional to the temperature difference between the soil and the coolant. Figure 3 shows the graphs of the temperature difference between the incoming liquid and the adjacent soil for model A and for model F.

\begin{figure}[h]
\centering
\includegraphics{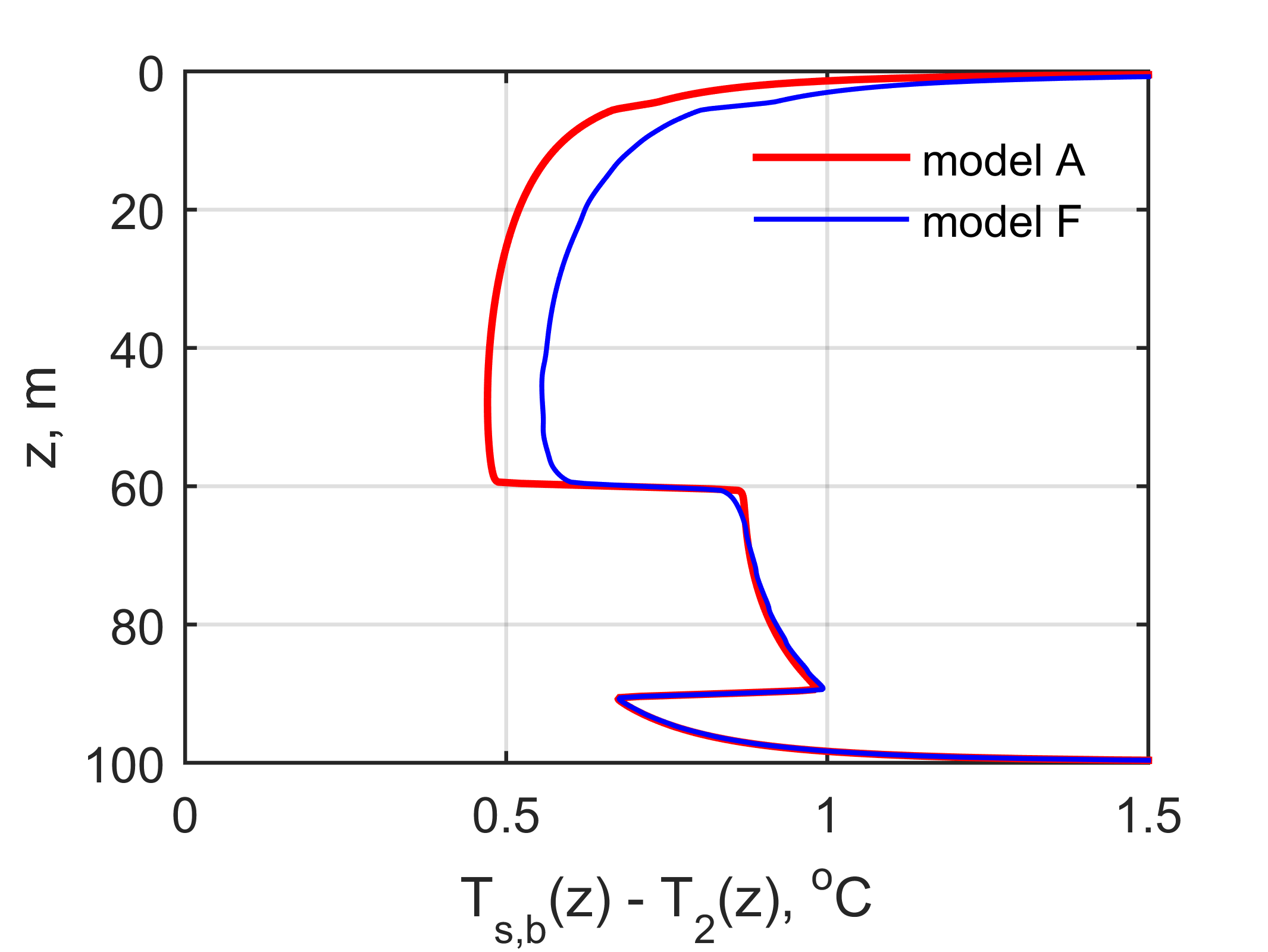}
\caption{The temperature difference between the adjacent soil and the incoming fluid, $T_{s,b}(z)-T_2(z)$,  at $T_{in} = -10^o$C, $v_f$ = 0.3 kg/s.}
\label{statb2}
\end{figure}

The color map of the temperature field in the vicinity of the borehole is shown in Fig. 4, on the left -- for model A and on the right -- for model F. The numbers near the temperature level lines show the temperature value on the corresponding line. It can be seen that the soil temperature in the model with freezing (F) is lower than the soil temperature in the model without freezing. This is understandable, since it follows from Figures 2 and 3 that the amount of heat extracted from the soil per unit time in model F is greater than in model A.

\begin{figure}[h]
\centering
\includegraphics{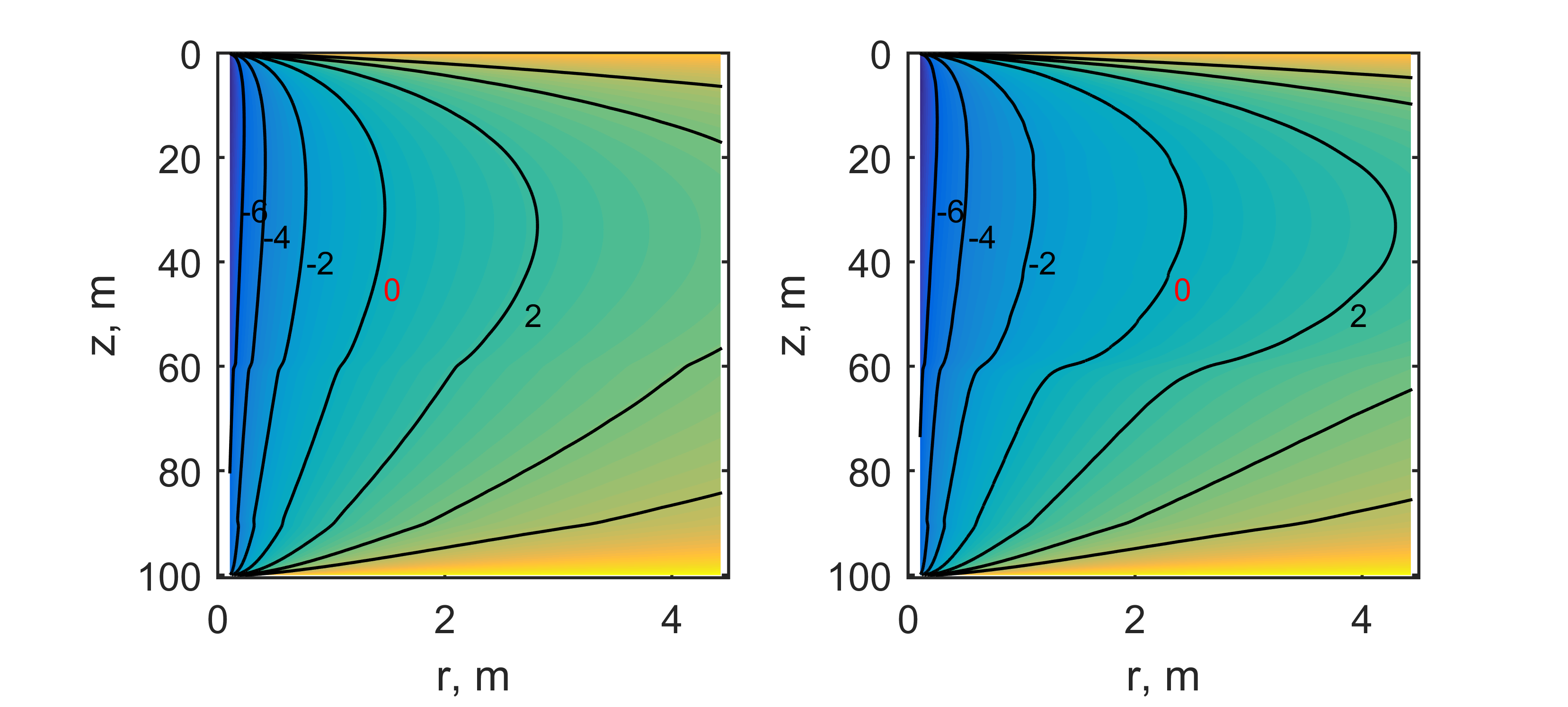}
\caption{Color map of soil temperature in stationary mode at $T_{in}$ = -10$^\circ$C, $v_f$ = 0.3 kg/s. The number at the level line indicates the temperature in $^\circ$C. Left -- model A, right -- model F.}
\label{statb3}
\end{figure}

However, it also follows from Fig. 2 that the temperature of the soil adjacent to the borehole in model F is higher than in model A. This apparent contradiction is clarified by Fig. 5, which shows plots of soil temperature depending on the radius for $z$ = 50 m (the middle of the exchanger). The temperature graphs intersect at a distance of approximately 10 cm from the borehole wall. To the left of the intersection point, the temperature in model F is higher than in model A, and to the right, it is vice versa.

\begin{figure}[h]
\centering
\includegraphics{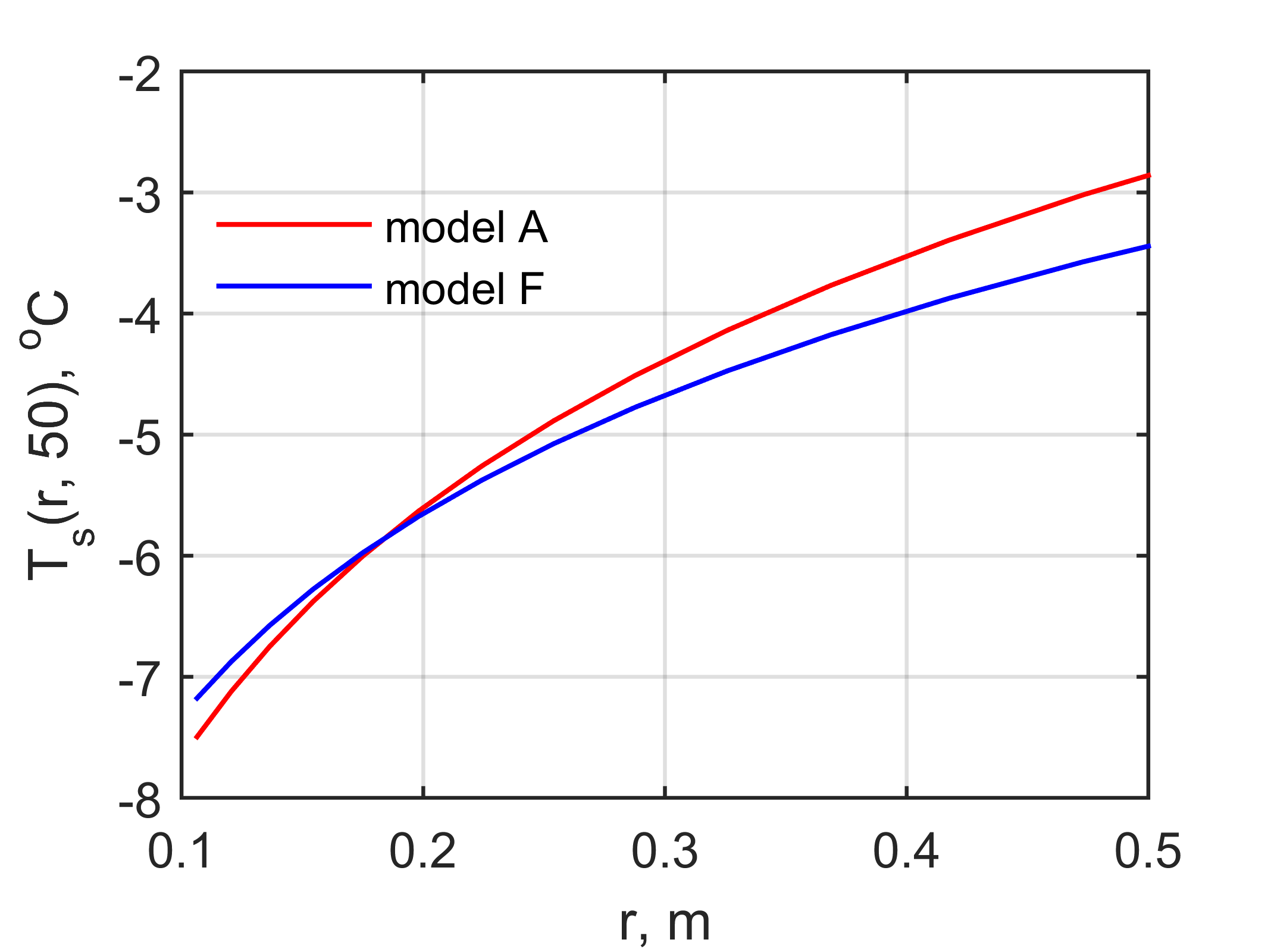}
\caption{Soil temperature versus radius at a depth of 50 m in models A and F. $T_{in}$ = -10$^\circ$C, $v_f$ = 0.3 kg/s.}
\label{stata4}
\end{figure}

In stationary mode, it is convenient to evaluate the sensitivity of various characteristics of the heat exchanger to changes in its parameters. In particular, let us estimate the sensitivity of the stationary heat extraction rate, $q_{st}$. In a linear approximation, the change in $q_{st}$ with a change in the parameter $p$ is calculated by the formula
\[\Delta q_{st} \approx \frac{\partial q_{st}}{\partial p}\Delta p,\]
where $p$ is one of the GHE parameters listed in Table 1. We estimate the partial derivatives numerically using the central finite difference by calculating $q_{st}$ at $p = (1\pm 0.01)p_0$, where $p_0$ is the value of parameter $p$ specified in the table. The results of calculation are presented in Table 3. 

\begin{table}[h]
\caption{Parameters sensitivity}
\centering
\begin{tabular}{c r r r}
\hline
     & $T_{in}=+2^\circ$C & $T_{in}=-10^\circ$C, A & $T_{in}=-10^\circ$C, F \\
 \hline
$r_{11}$ & $-$1.732e+04 & $-$3.758e+04 & $-$4.582e+04 \\
$r_{12}$ & 1.568e+04 & 3.401e+04 & 4.148e+04 \\
$r_{21}$ & 7.572e+01 & 1.672e+02 & 2.825e+02 \\
$r_{22}$ & 1.868e+03 &  4.125e+03 & 5.444e+03 \\
$r_b$ &  2.985e+02 & 7.039e+02 & 1.517e+01 \\
\hline
$\lambda_1$ &  $-$1.231e+02 & $-$3.563e+01  & $-$3.293e+02 \\ 
$\lambda_2$ & 1.092e$-$02 & 2.412e$-$02 & 2.833e$-$02 \\ 
$\lambda_g$ & 2.223e+01 & 4.909e+01 & 6.518e+01 \\ 
$\alpha_{11}$ & $-$6.808e$-$03 & $-$1.477e$-$02 & $-$1.786e$-$02 \\ 
$\alpha_{12}$ & $-$2.533e$-$02 & $-$5.495e$-$02 & $-$6.507e$-$02 \\ 
$\alpha_{21}$ & 1.121e$-$02 & 2.475e$-$02 & 4.180e$-$02 \\ 
\hline
$c_f$ & 3.964e$-$02 & 8.725e$-$02 & 1.094e$-$01 \\
$\rho_f$ & 1.137e$-$12 & $-$1.439e$-$11 & $-$1.105e$-$11 \\
$v_f$ & 5.946e+02 & 1.309e+03 & 1.642e+03 \\
$T_{in}$ & $-$1.124e+02 & $-$1.242e+02 & $-$1.382e+02 \\
\hline
\end{tabular}
\end{table}

Table 3 shows estimates of partial derivatives for both positive, $T_{in}=2^\circ$C, and negative, $T_{in}=-10^\circ$C, inlet temperatures. Moreover, for a negative temperature, the derivatives were calculated in both models A and F. In all calculations for the heat extraction rate $q_{st}$, $r_{11}$ and $r_{12}$ turned out to be the most sensitive parameters. The derivatives with respect to them have different signs. The derivative with respect to $r_{11}$ is negative. An increase in $r_{11}$ leads to a decrease in the thermal resistance of the inner pipe wall and to a decrease in the linear velocity of the fluid in the inner pipe. Therefore, the fluid cools down more strongly when moving through the inner pipe, which leads to a decrease in $q_{st}$. The derivative with respect to $r_{12}$ is positive. An increase in $r_{12}$ leads to an increase in thermal resistance between the fluids in the outer and inner pipes, so the outgoing fluid cools less and $q_{st}$ increases. Similar conclusions about the effect of  inlet temperature and  inner and  outer pipes diameters on the thermal performance of a  coaxial GHE were reported in \cite{Liu_2019}  for a deep  GHE.

Note that under given conditions, all derivatives of $q_{st}$ with respect to the heat exchanger parameters have the same signs and almost all of them are of the same order of magnitude. The exceptions are the derivatives with respect to $r_b$ and $\lambda_1$. The derivative with respect to $r_b$ in model A is 46 times greater than in model F, and the derivative with respect to $\lambda_1$ in model A is approximately 10 times smaller in absolute value than in model F.

The stationary solution is easy to calculate and provides useful information about the heat extraction process. However, with a significant difference between the average soil temperature (11.5$^\circ$C in our example) and the inlet temperature $T_{in}$, the stationary solution turns out to be practically unattainable due to the large relaxation time. Figure 6 gives an idea of the relaxation time for the assumed conditions. 
\begin{figure}[h]
\centering
\includegraphics{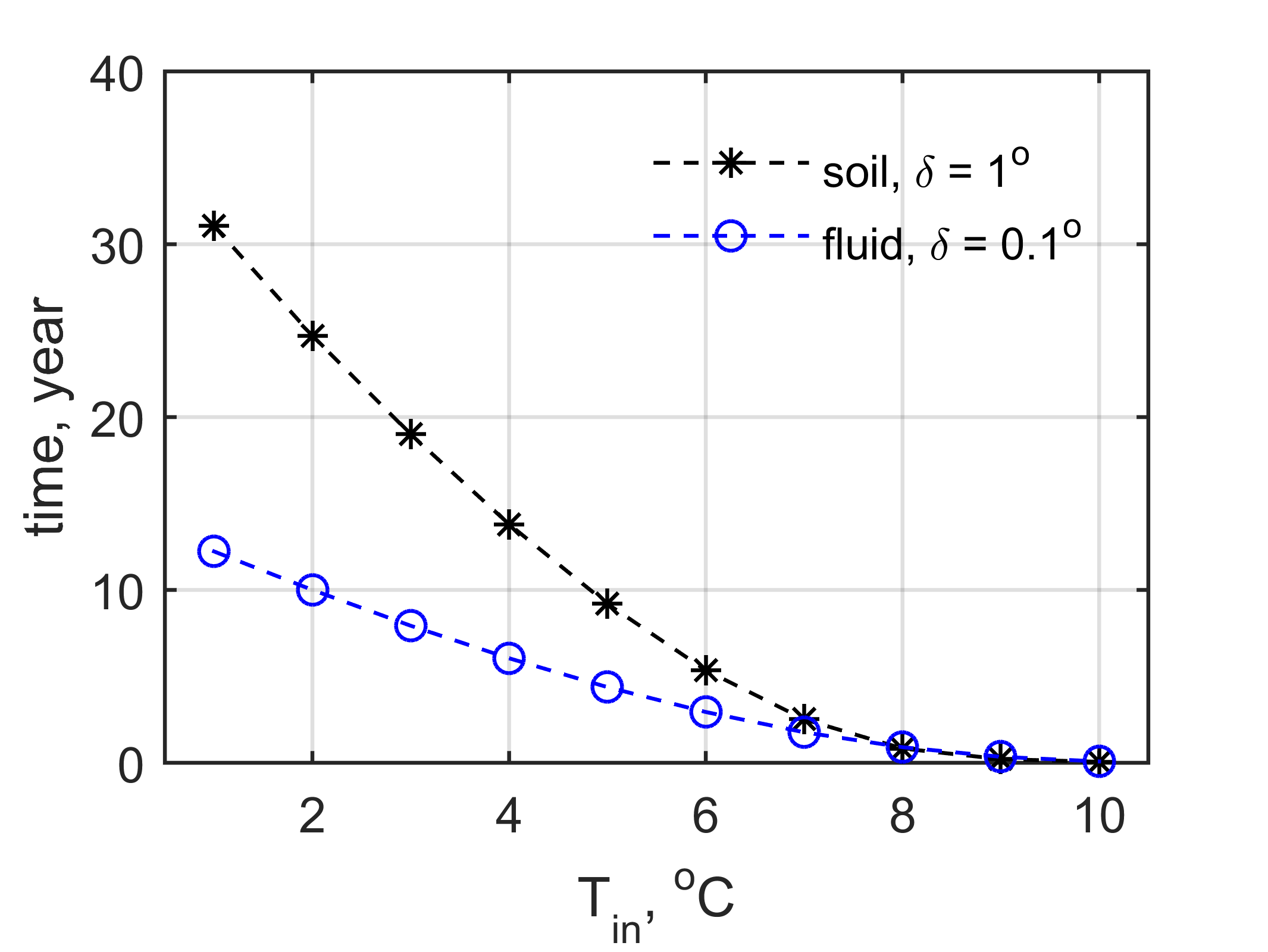}
\caption{Estimation of the duration of the relaxation time of soil and fluid depending on the inlet temperature.}
\label{trel}
\end{figure}
The relaxation time increases rapidly as the difference between the natural ground temperature and the inlet fluid temperature increases. Moreover, the temperature of the fluid in the heat exchanger approaches the equilibrium temperature much faster than the temperature of the soil. We estimate the time of approaching the soil temperature to equilibrium by the time $t_*$ at which the inequality $\max_{r,z}|T(t,r,z)-T_{st}|<\delta$ is satisfied for the first time, where $T$ is the solution of (\ref{mat}) and $\delta=1^\circ$C. For the fluid, we estimate the time $t_\circ$ at which the inequality $T_{out}(t)-T_{out}^{st}<\delta$ begins to hold for $\delta=0.1^\circ$C. The graphs of $t_*$ and $t_\circ$ in dependence on $T_{in}$ are shown in Figure 6.

\subsection{Transient mode}

Let us now return to system (\ref{mat}) and present the time-dependent solution of the system immediately after the start of working. Here and below, all calculations are performed with the initial conditions (\ref{tf0}), (\ref{ts0}).

Figure 7 shows graphs of fluid temperature, $T_{out}(t)$, at the outlet of the heat exchanger for the first five days of operation at $T_{in}=-10^\circ$C for both models A and F.  \begin{figure}[h]
\centering
\includegraphics{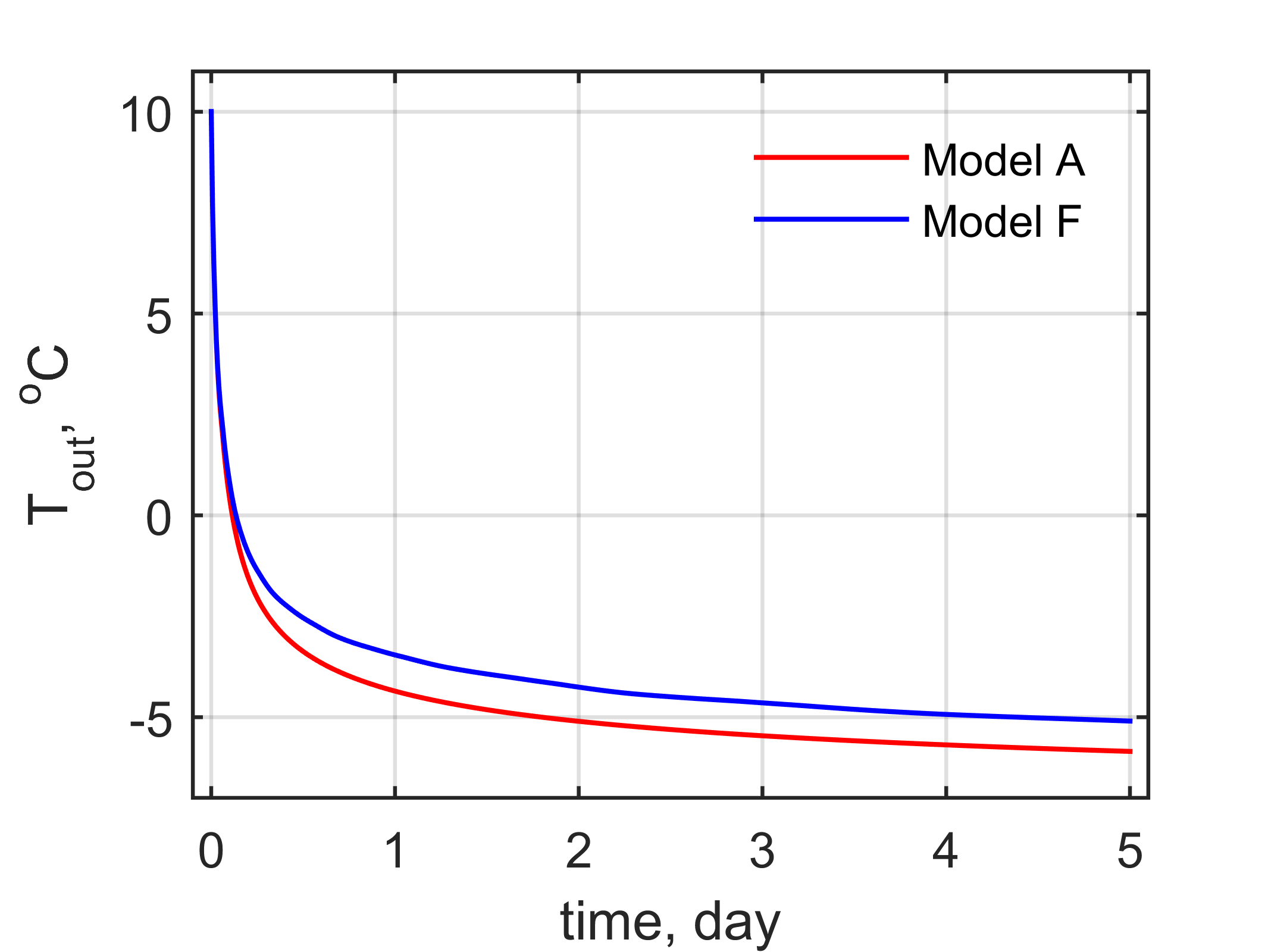}
\caption{Outlet temperature $T_{out}(t)$ in model A and F. $T_{in} = -10^\circ$C, $v_f$ = 0.3 kg/s.}
\label{tout7}
\end{figure}

During the first day of operation, the temperature $T_{out}(t)$ drops rapidly. By the end of the fifth day, the temperature drop noticeably slows down, although it is still very far from the equilibrium temperature, $-8.2^\circ$C for model A and $-8.0^\circ$C for model F (see Fig. 2). The convergence of the solution of both models to the stationary solution is very slow. From Figure 6, we can conclude that the relaxation time for $T_{in} = -10^\circ$C must be very large and thermal equilibrium in this example is practically unattainable.

By the end of the fifth day the outlet temperature in model A equals to $-5.85^\circ$C and in model F it equals $-5.10^\circ$C. The heat extraction rate $q_{ex}$ is proportional to the difference $T_{out}-T_{in}$. Thus, under given conditions, by the end of the fifth day of work, $q_{ex}$ calculated using model F is 18\% more than $q_{ex}$ calculated using model A.  

The evolution of soil temperature near the borehole in the beginning five days of operation is shown in Figures 8 and 9. These figures show temperature versus time graphs at nine spatial grid nodes with coordinates $z$ = 0.25 m and $r$ = 0.106, 0.120, 0.136, 0.154, 0.175, 0.198, 0.224, 0.254, 0.288 m, from bottom to top (borehole radius $r_d$ = 0.1 m). These nodes are located in horizontal plane just near the top of the exchanger ($z$ = 0). 

Figure 8 shows plots of soil temperature in model A, obtained without taking into account moisture freezing. Here the temperature changes smoothly, 

\begin{figure}[h]
\centering
\includegraphics{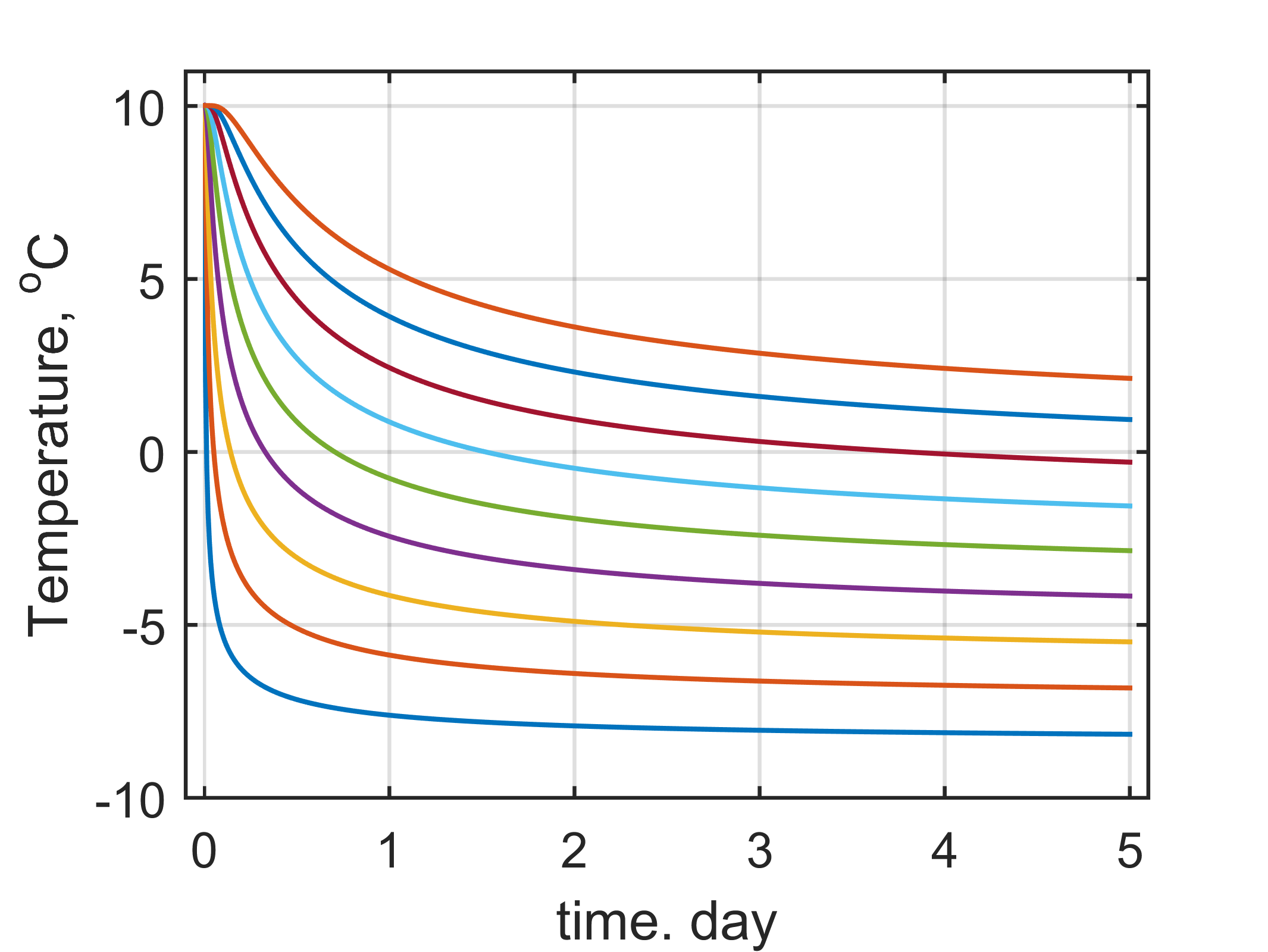}
\caption{Soil temperature in model A at 9 grid nodes (see text). $T_{in}$ = -10$^\circ$C, $v_f$ = 0.3 kg/s.}
\label{tout7}
\end{figure}

\begin{figure}[h]
\centering
\includegraphics{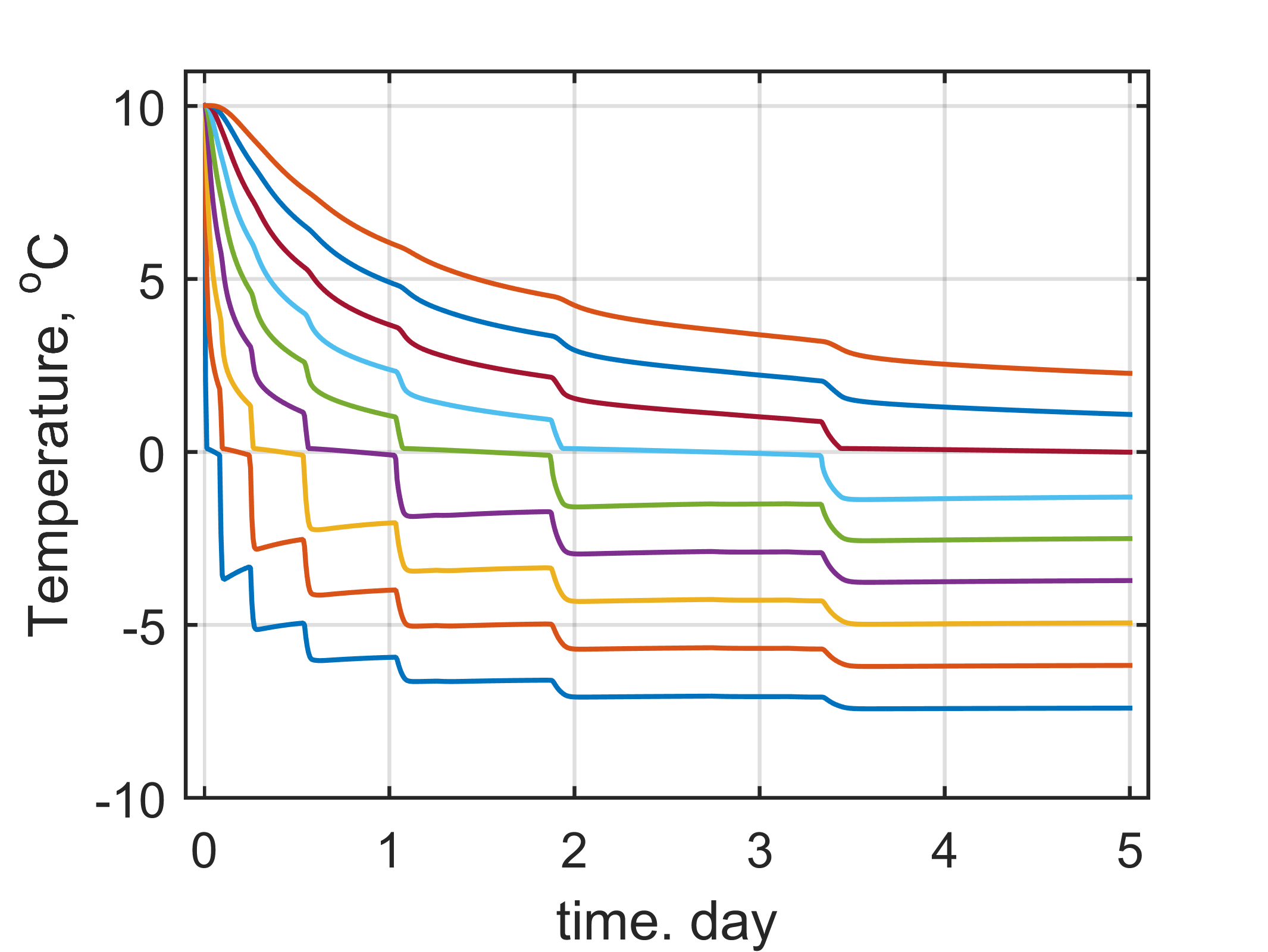}
\caption{Soil temperature in model F at the same 9 grid nodes as in Fig. 8. $T_{in}$ = -10$^\circ$C, $v_f$ = 0.3 kg/s.}
\label{tout8}
\end{figure}

\clearpage
\noindent
decreasing rapidly on the first day and slowing down noticeably at the end of the fifth day. Note that the temperature curves are almost equidistant. This fact is a consequence of the exponential grid in the radial direction.

A different view have the soil temperature curves at the same grid nodes in Figure 9, calculated using the model F. When the temperature in the node and in the corresponding cell of the spatial grid decreases to $T_f-\Delta_f$, moisture begins to freeze in the cell. During freezing, latent heat $L$ is released, which maintains the temperature in the interval ($T_f-\Delta_f,\, T_f+\Delta_f)$ in the grid cell and slows down the temperature change in neighboring cells in the radial direction. This phenomenon is known as the zero-curtain effect \cite{Outcalt_1990}.

The farther the node is from the borehole, the larger the volume of the cell and the more latent heat is released. Therefore, with the distance from the borehole, the freezing time of moisture in the cell increases. Temperature stabilization in the cell for the time of moisture freezing slows down the temperature change not only in the cells with a negative temperature located closer to the borehole wall, but also in several adjacent cells with a positive temperature, which lie farther from the borehole in the same horizontal plane. The release of the latent heat of freezing increases the heat flux to the heat exchanger and increases the temperature of the fluid in the outer pipe, which is confirmed by the graphs in Figure 7.

Obviously, the stepped form of the curves is a consequence of the discretization of the problem. With refinement of the spatial grid, the number of steps will increase, their height will decrease, and the curves will be smoothed out.

Figure 10 shows color maps of soil temperature near the borehole -- in the area of the most rapid temperature change with distance from the borehole. The general slope of the level lines to the right is due to the geothermal gradient. Soil temperature at $z$ = 100 (13$^\circ$C) is three degrees higher than at $z$ = 0. The breaks in the temperature level lines on the left map (model A) correspond to the boundary between the soil layers with different physical properties. These kinks are also visible on the right map (model F), but here one can also see much more kinks associated with temperature stabilization when moisture freezes. Level lines corresponding to the same temperature in model A are located farther from the borehole than in model F. Therefore, at the end of the fifth day, the amount of heat in the soil calculated by model A is less than in model F. There is no contradiction here with Fig. 4, because in Fig. 4 shows the stationary (final) temperature distribution in the soil, and Fig. 10 shows the unsteady  transitional distribution.

\begin{figure}[h]
\centering
\includegraphics{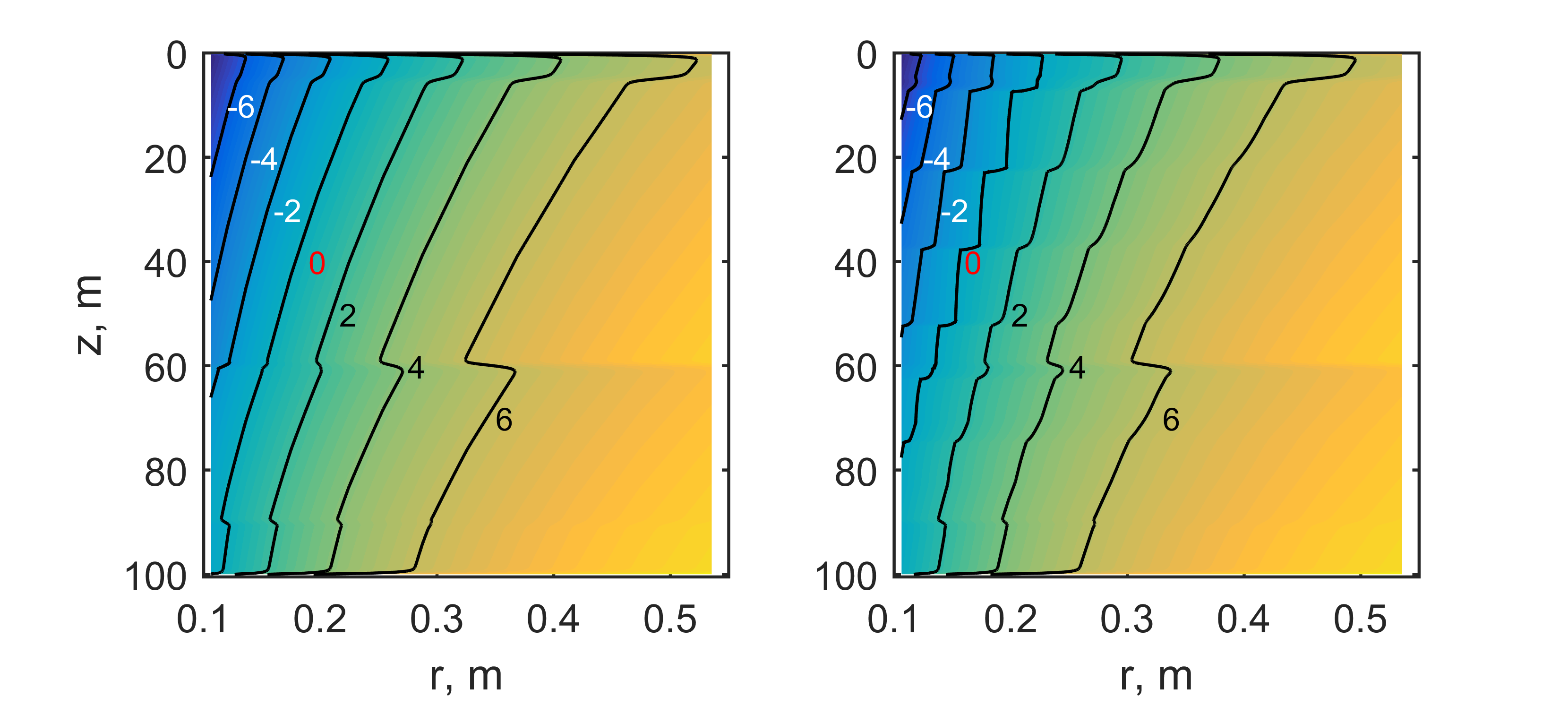}
\caption{Color map of soil temperature near the well at the end of the fifth day of heat exchanger operation. Left -- model A, right -- model F. $T_{in}$ = -10$^o$C, $v_f$ = 0.3 kg/s.}
\label{tout9}
\end{figure}

\subsection{Controlled mode}

And, finally, we give an example of a controlled mode of operation of the heat exchanger. Suppose that during the heating season (October 1 - April 31) every day the heat exchanger must eject from the soil the amount of heat $Q_d$, proportional to the difference between the average daily ambient temperature per day $d$, $T_a^d$, and room temperature $T_r=20^\circ$C,
\[Q_d=k_q(T_a^d-T_r),\]
where $k_q$ [J/K] is a constant coefficient. 

For definiteness, let's take the Moscow region. The graph of the average daily air temperature in Moscow is shown in Fig. 11.

\begin{figure}[h]
\centering
\includegraphics{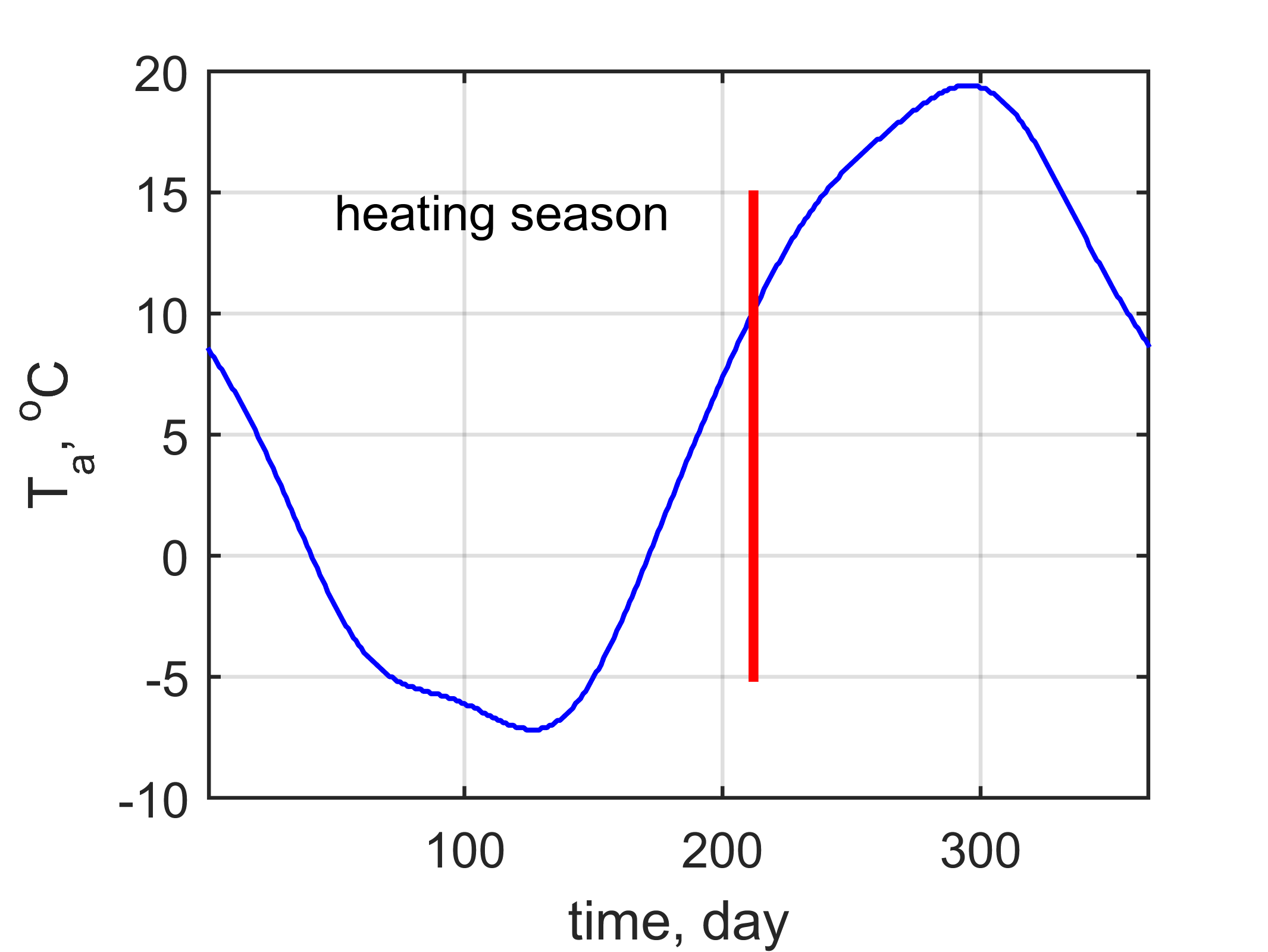}
\caption{Average daily temperature in Moscow. First date -- October 1st.}
\label{tTA}
\end{figure}

Let the temperature of the fluid at the inlet to the heat exchanger $T_{in}$ during the $d$-th day be constant and equal to $T_{in}^d$, then the heat exchanger takes away on the $d$-th day the amount of heat $Q_{ex}^d$ equal to
\[Q_{ex}^d(T_{in}^d)=c_fv_f\int_d{(T_{out}(t)-T_{in}^d)\,dt},\]
where $T_{out}(t)$ is obtained from the solution of Eq. (\ref{mat}) with $T_{in}=T_{in}^d$ and the temperature field at the end of the previous day as the initial condition. The integral is taken over $d$-th day. To obtain the required amount of heat on each day of the heating season, it is necessary to solve the following chain of equations for $T_{in}^d$,
\begin{equation}
\label{eqq}
Q_{ex}^d(T_{in}^d)=Q_d,\; d = 1,2,...
\end{equation}

Figure 12 shows the results of calculations of the heating period, performed according to the F and A models for two values of the coefficient $k_q$, 10$^4$ and $2\times 10^4$. The top panel shows plots of extracted heat $Q_{ex}(t)$. On the scale of the Figure, the graphs of $Q_{ex}(t)$ and the required heat $Q_d(t)$ coincide. The lower panel shows graphs of the inlet temperature of the fluid $T_{in}$ corresponding to the required amount of heat $Q_d$, calculated by model A and model F.

\begin{figure}[h]
\centering
\includegraphics{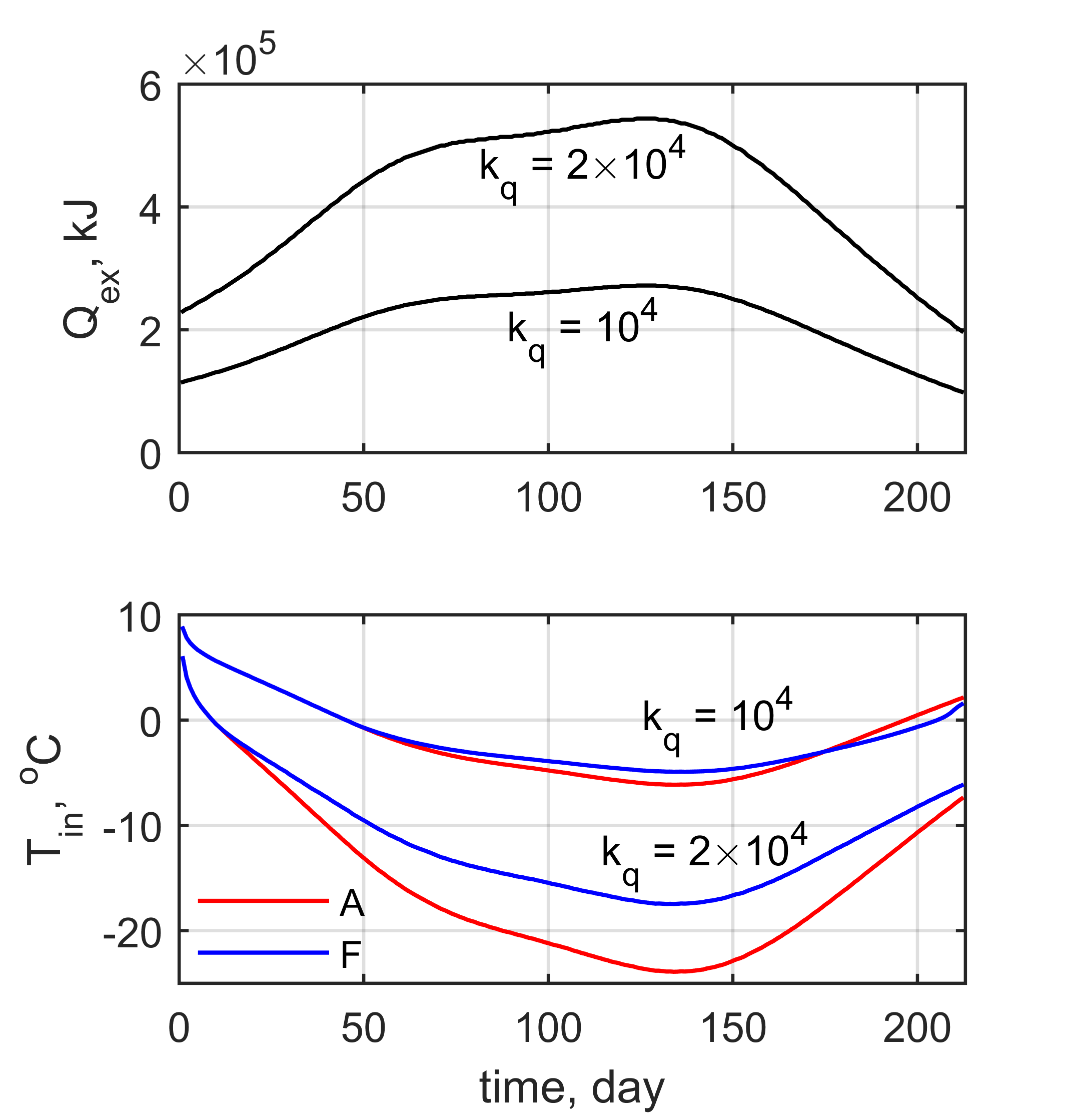}
\caption{Extracted heat (top) and inlet temperature (bottom) in controlled mode.}
\label{tTA}
\end{figure}

From Fig. 12 it follows that for both considered values of $k_q$, negative fluid temperatures are necessary to satisfy the required amount of heat extracted from the soil. Note that if at $k_q=10^4$ the difference in the inlet temperatures of the fluid in the heat exchanger, calculated according to models A and F, is small, then at $k_q=2\times 10^4$ the maximum difference reaches 37 percent: $-23.9^\circ$C in model A and $-17.4^\circ$C in model F.

The numerical solution of equation (\ref{eqq}) for the $d$-th day was obtained by a method similar to the method of chords for solving nonlinear equations. First, the amount of heat $Q_1$ extracted at the inlet temperature $T_1=T_{in}^{d-1}$ is calculated. Then the amount of heat $Q_2$ extracted from the soil at the inlet temperature $T_2 = T_1 +\delta$, if $Q_1>Q_d$, or at $T_2=T_1-\delta$, if $Q_1< Q_d$ is calculated. Then 
\[T_3=T_1+\frac{Q_d-Q_1}{Q_2-Q_1}(T_2-T_1)\]
was assigned and the amount of heat $Q_3$ extracted at the inlet temperature $T_3$ was calculated. The described iterations stopped when the condition
\[|Q_3-Q_d|/Q_d < 0.001.\]
Note that in all our calculations, this condition was already satisfied at the first iteration.

The temperature field in the soil in the vicinity of the borehole at the end of the heating period is shown in Fig. 13. The temperature is calculated according to the F model, i.e. taking into account the freezing/thawing of moisture, on the left -- for $k_q=10^4$, on the right -- for $k_q=2\times 10^4$. At $k_q=10^4$, the ice has completely melted by the end of the heating period, and the ground temperature is everywhere positive. At $k_q=2\times 10^4$, a significant part of the moisture in the soil remains frozen. In the upper part of the borehole, the maximum ice boundary approximately runs at a distance of 0.78 m from the borehole axis.

\begin{figure}[h]
\centering
\includegraphics{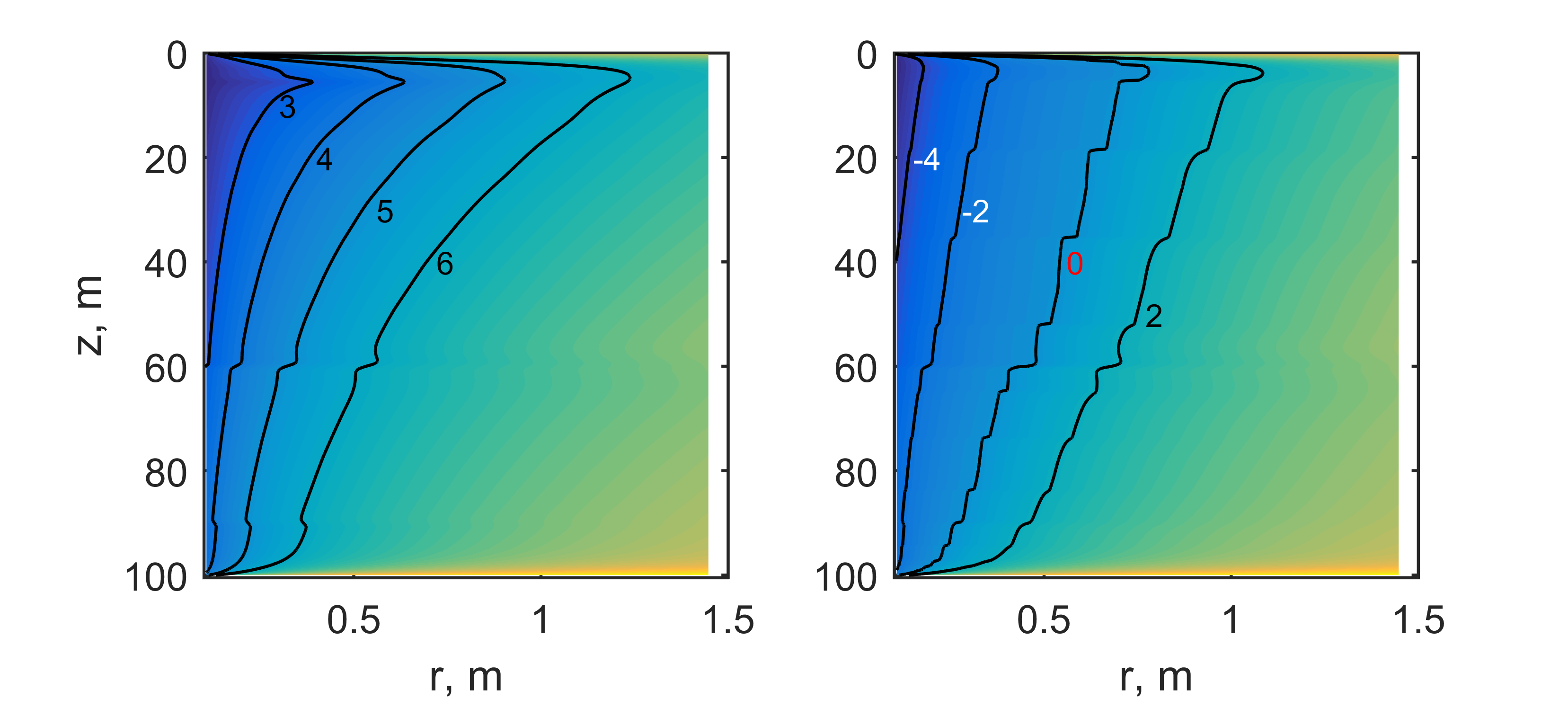}
\caption{Soil temperature at the end of heating season (model F). Left -- $k_q=10^4$ J/K, right -- $k_q=2\times 10^4$ J/K.}
\label{tTA}
\end{figure}

Thus, at the end of the heating season, in both considered cases, the soil temperature near the borehole is significantly lower than the undisturbed soil temperature. Therefore, the question arises of restoring the soil temperature by the beginning of the next heating season. In this paper, we do not consider this issue, but note that the natural temperature recovery due to the influence of boundary conditions cannot be correctly described by our model. Our model does not take into account the convective heat transfer in fluids in the vertical direction. When fluid moves in a working heat exchanger, this transfer can be ignored. However, in a stagnant fluid in the summer, it can have a significant effect on the restoration of the soil temperature near the borehole, in particular, on the melting of adjacent ice.

\section{Conclusion}

The paper proposes a new mathematical model of heat transfer in a shallow vertical coaxial ground heat exchanger and in the surrounding ground. The model takes into account soil heterogeneity and geothermal gradient. At a negative temperature of the heat-carrying fluid, the process of freezing-thawing of ground moisture is taken into account. The water-ice phase transition is modeled using the apparent heat capacity method. The numerical implementation of the model was performed using the finite difference method on a non-homogeneous spatial grid under the assumption of axial symmetry of the problem. The computer implementation of the model is made in the Matlab environment on a personal computer, which makes the model accessible to a wide range of specialists.

In the considered examples, three modes of operation of the heat exchanger are modeled: stationary, transitional and controlled. In the calculations, the main attention was paid to demonstrating the differences in the results obtained with and without taking into account the water-ice phase transition in ground moisture. From the above examples, it follows that the greatest difference is noted in the controlled extraction of heat from the soil during the heating season.

The proposed model will be useful in the design of heat pump installations for heating and cooling buildings if it is necessary to work with negative temperatures of the coolant, taking into account the structural features and composition of the soil. To simulate the release of heat into the soil, it is enough to change the direction of fluid movement in the model. The model can also be used to optimize the parameters of the heat exchanger, taking into account the specific conditions of its operation.

\section*{Disclosure of Potential Conflicts of Interest}

The Authors declare that there is no conflict of interest.

\section*{Acknowledgment}

\end{document}